# Impulsive pattern recognition of a myoelectric hand via Dynamic Time Warping

Kadılar M.C., Graduate Student, Toptaş E., Asst. Prof., Akgün G., Asst. Prof.

*Abstract*—Although myoelectric prosthetic hands provide amputees with intuitive control, their reliance on many EMG sensors limits accessibility and makes them complex and expensive. To address this problem, this work presents a different perspective that makes use of a single EMG sensor and brief impulse signals in conjunction with Dynamic Time Warping (DTW) for accurate pattern detection. Conventional techniques rely on real-time data from multiple sensors, which can be costly and bulky. The method presents high accuracy while lowering hardware complexity and expense. A DTW-based system that reliably identifies muscle activation patterns from short EMG signals was created and tested. Results show that this single-sensor approach obtained an accuracy rate of 92%, which is similar to that of conventional multi-sensor systems. This research provides a more straightforward and economical approach that can be used to obtain enhanced myoelectric control. These findings provide a different perspective on more easily accessible and user-friendly prosthetic devices, which will be especially helpful in disaster-affected areas where quick deployment is essential. Future improvements would investigate this system's dependability over time and wider implementations in real situations, to take prosthetic technology one step further.

## I. INTRODUCTION

Originally made for cosmetic reasons in the fifteenth century B.C., prosthetic limbs in Egypt have been researched and improved by scientists and manufacturers since then. Over time, prosthetics have included functional elements including metal bits that could be moved to grip, lock, and bend objects. By using electromyography (EMG) signals—biological impulses produced by muscle contraction—modern prostheses demonstrate how far the field of prosthetics has come and allow more natural control. Particularly for those who are basic and offer good feedback, the development of prosthetic technology will determine how well people who have lost limbs live.

This requirement becomes more critical following natural disasters such as earthquakes when limb injuries and eventual amputations are frequent occurrences.

The significance of developing prosthetic technology is shown by the happenings of late. One orthopaedic clinic, for example, handled 560 patients during the 2023 Kahramanmaraş earthquake; 31 of them needed amputation. Likewise, of the 120 patients treated for soft tissue injuries to the upper extremities in a plastic surgery department [5], 46.2% needed an amputation, most of which were for the upper extremities. These tragedies demonstrate how urgently advanced prosthetic solutions are needed. Furthermore, [6] estimates indicate that one in 190 Americans now suffers from limb loss; if present trends persist, that figure could double by 2050.

Notwithstanding their advancements, myoelectric prostheses still present problems [7]. Uneven control and harsh, unpleasant motions are common complaints of myoelectric hand users and physical therapists. These challenges can cause myoelectric prosthesis rejection rates of up to 23%. These problems regarding the challenges of day to day life use of prosthetic hand users are analyzed and researched by many [7, 8, 9, 10]. Atasoy et al. designed a 24 degress of freedom prosthetic hand which categorizes biological data to various grips using multiple sensors [8]. Another group of researchers, Wattanasiri et. al. used a single motor and a multifunctional grip that would be applicable for many instantces of challeneges of daily life of a user [9]. Though their design was inappropriate for high-torque conditions, Ahmed et al. [10] considered substituting pressure sensors for EMG sensors for muscular signals. Ismail et al. [11] took a reasonable approach by providing a user interface to record muscle data for a five-fingered prosthetic hand.

This work offers a more straightforward method by reducing the sensor requirement with a pattern recognition algorithm over a single sensor. Utilizing simple signals from a single sensor to create multiple patterns improves on what the other studies build upon. The results show that prosthetic hands can be controlled efficiently with a single-sensor, impulse-based technique that can match conventional multi-sensor systems. This development streamlines prosthesis design, which could lower costs and increase user accessibility. Even with little sensor data, the system can reliably recognize and categorize muscle activation patterns by utilizing Dynamic Time Warping (DTW) for pattern recognition. The study provides a different perspective on the systems that have been developing over the years by experimenting with single-sensor pattern controls. It demonstrates the viability of using a single EMG sensor for sophisticated myoelectric control.

## II. METHOD

*A. Required Mechatronical Setup*

Workflow of this study is shown below, in figure 1.

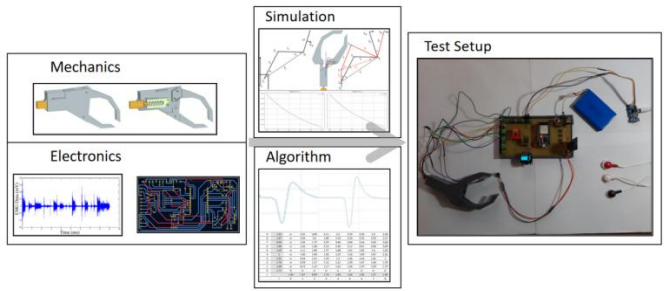

**Fig. 1.** Study workflow.

The study seeks to test and enhance the control options of a myoelectric-controlled prosthetic hand. To achieve this, a test setup had to be created to be experimented on, composed of mechanical and electronic systems. The basic mechanical system is composed of an Encoder, a DC motor, a worm gear, and a sliding piece that is connected to two different pins that move the fingers of the prosthetic, as shown in Figure 2 below. The encoder provides precise control and prevents the prosthetic hand to be open/closing beyond its mechanical limits by measuring each rotation of the DC motor, while worm gear provides the mechanical advantage needed for a prosthetic hand to be able to grip objects for testing. These designs are printed with a 3D printer using PLA filaments. This mechanical system goes through simulation to predict the resultant product's kinematics, such as the speed at the tip of the fingers, to create the optimal mechatronic design.

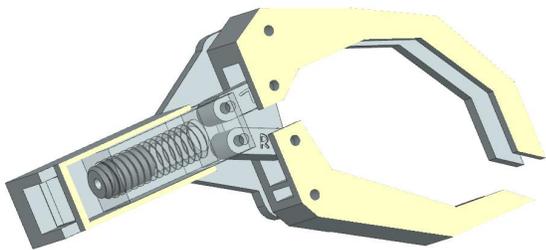

**Fig. 2.** Internal view of the mechanical system

The basic electronics system is composed of an EMG sensor and a control board. The three EMG probes are connected to the forearm muscle palmaris longus as shown in Figures 3 and 4 below. measures the biological signals as shown in Figure 5 below, and outputs a voltage difference for the control board to process through an algorithm called Dynamic Time Warping (DTW).

The EMG data will be gathered by means of the EMG probes fastened to a person's muscle, as shown in Figure 3 below. After being filtered, the collected data is sent to an ESP32 for additional processing. EMGs typically use three probes to collect data, as shown. They are as follows: live (red), which produces the signals; neutral (black), which serves as the live probe's reference point; and ground (white), which filters the signals.

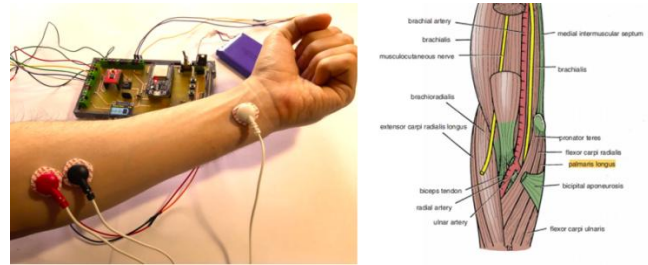

**Fig. 3&4.** Connection of EMG probes, as shown on the right side, to the muscle 'palmaris longus' [12].

The EMG probes in Figure 3 are attached to the arm at the location shown in Figure 4 at the right side of it. Below in Figure 5 actual EMG data is displayed; the signal variation reflects muscle contractions that the DTW algorithm filtered. Which gets processed via DTW after filtering. Every pulse is a muscle activity read.

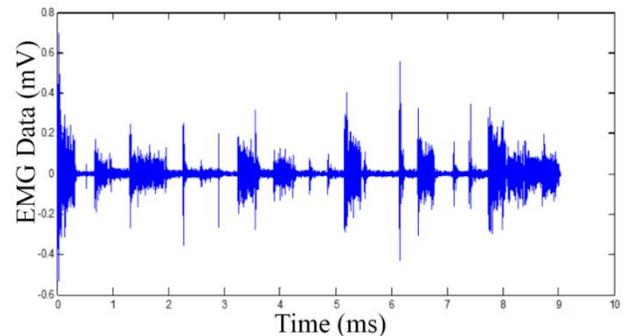

**Fig. 5.** An unfiltered data from surface EMG, retreived from [13]

*B. DTW What and How*

The EMG SpeedStudio, with its built-in filtering system, will later filter this data. The filtering system produces more dependable data for further processing by eliminating the electrical noise such as the nearby equipment and biological noise.

TABLE I
EXAMPLE DTW MATRIX WHICH USES THE DATA FROM FIGURE 5.

| | | | | | | | | | |
|---|---|---|---|---|---|---|---|---|---|
| 9 | 1.52 | ∞ | 2.31 | 3.09 | 2.11 | 2.2 | 2.29 | 2.28 | 2.3 | 2.28 |
| 8 | 1.57 | ∞ | 2.26 | 2.6 | 1.89 | 2.15 | 2.24 | 2.25 | 2.25 | 2.27 |
| 7 | 0.93 | ∞ | 2.26 | 1.72 | 2.29 | 3.06 | 2.84 | 2.64 | 2.65 | 2.68 |
| 6 | 1.06 | ∞ | 1.62 | 1.48 | 2.16 | 2.45 | 2.11 | 2.01 | 2.06 | 2.09 |
| 5 | 1.52 | ∞ | 1.11 | 1.89 | 1.77 | 1.68 | 1.51 | 1.55 | 1.6 | 1.63 |
| 4 | 2 | ∞ | 1.06 | 1.94 | 1.55 | 1.37 | 1.54 | 1.89 | 2.07 | 2.26 |
| 3 | 1.91 | ∞ | 0.63 | 1.51 | 1.29 | 1.2 | 1.45 | 1.64 | 1.81 | 2 |
| 2 | 1.74 | ∞ | 0.29 | 1.17 | 1.12 | 1.21 | 1.29 | 1.47 | 1.64 | 1.78 |
| 1 | 1.69 | ∞ | 0.12 | 1.12 | 1.17 | 1.31 | 1.34 | 1.47 | 1.59 | 1.73 |
| 0 | 1.71 | 0 | ∞ | ∞ | ∞ | ∞ | ∞ | ∞ | ∞ | ∞ |
| j | | 1.56 | 1.57 | 0.69 | 1.74 | 1.83 | 1.66 | 1.56 | 1.57 | 1.55 |
| | i | 0 | 1 | 2 | 3 | 4 | 5 | 6 | 7 | 8 |

Filtering and amplifying EMG signals will help to make them more consistent and reliable, as Abdhul et al. [3] describe. Following the filtering procedure, EMG signals go through an algorithm called Dynamic Time Warping (DTW). An example EMG data set in which the signal variation reveals muscle contractions filtered with the DTW algorithm is shown in Figure 5. Described in [14], the DTW is a method that provides a similarity value for two data graphs by comparing two data point sequences. Often employed in time series analysis when timelines are shifted back or forth, it makes traditional distance measurements like Euclidean distance insufficient. DTW sidesteps the problems caused by isalignment or event variability in real time influenced by signal timing. By means of comparing muscle contraction graphs over time, DTW is used in this work to generate a similarity value. The comparison of the biological signals from the contracted, relaxed and resting muscles produces this similarity value. This approach seeks to produce a responsive and realistic prosthetic hand operation by contrasting the real time data with recorded data sequences.

In Figure 6 below, two graphs of data sequances are visible. Comparing these two with dynamic time warping algorithm would give a similarity value. The signals 'i' and 'j' are different in data point voltage on the left and time location on the bottom. The data set 'i' even has one more data point on time domain. As shown in Figure 6 and Equation (1), if DTW algorithm is applied a similarity value will be generated. This value is placed and shown in Table 1 above, 0.29. The Equation (1), the initial value of $D_{i,j} = d(x_i, y_j) + min$, is the distance between first and second data points between the graphs of 'i' and 'j'. As shown in Table 1 above, next points are the minimum of the bottom box, left box, and bottom left boxes. The values in these boxes are put through the algorithm and the DTW matrix starts to form. In Table 1, the DTW matrix continues to build with the Equation (1), until the top right of the table, the last box of values are filled, which will give the overall similarity value of both graphs, or data sequances, which in this case resulted in the value '2.67'. Further the values diverge from zero, the less similar the data sequences are. In reality, this algrorithm is run instantly in the microchip ESP32, constantly measuring the similarities of various datasets and biological signals, comparing them to the expected and recorded biological signals.

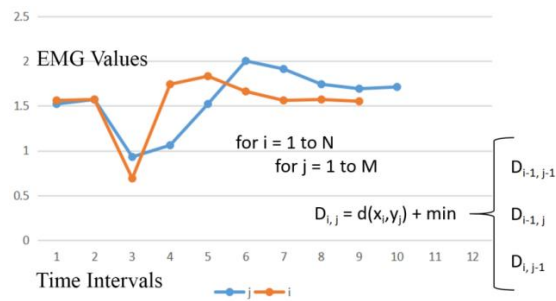

**Fig. 6.** Formula and example graphs for DTW.

$$D_{1,2} = |1.69 - 1.57| + min \begin{cases} D_{0,1} = \infty \\ D_{0,2} = \infty \\ D_{1,1} = 0.12 \end{cases} = 0.29 \quad (1)$$

A user interface, has assisted in generating the data gathered from the DTW similarity matrix. Hoshigawa et al. [13] addressed how myoelectric prosthetic hand control is constrained even with machine learning algorithms. On the other hand, this work used a user interface (UI) based calibration technique with a more stable algorithm as another control strategy.

Compared to DTW, the other pattern recognition algorithms have their own downsides. For example one of the most commonly used algorithms is called Support Vector Machine (SVM). SVM operates by locating the hyperplane in a high-dimensional space that best divides various data classes. SVM may be trained on labeled EMG data to categorize various muscle activation patterns into distinct hand movements, which is useful for myoelectric control such as DTW [16].

SVM requires way more training data and parameters to perform and although it can outperform DTW in some cases, SVM tends to work better in more stable datasets, any skip on data points or static can disrupt SVM outputs way more than DTW outputs. DTW is designed to filter out misalignments in data hence the name "warping" of the time domain.

III. RESULTS

*A. Patterns Recorded*

The raw signals coming from the EMG sensor are filtered and measured between 3.3V and 0V through the ESP32 shown in Figure 7 below. The left signal can be categorized as "Pattern 0" and the right signal can be categorized as "Pattern 1".

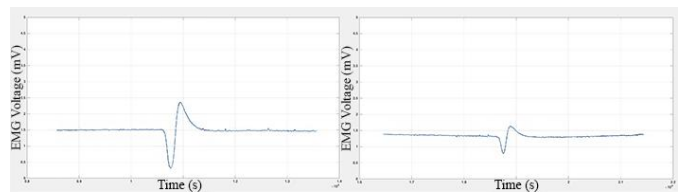

**Fig. 7.** EMG detected filtered muscle contraction patterns

The filtered signals are recorded in the SD Card with intervals of 20 milliseconds as shown in Figure 8 below.

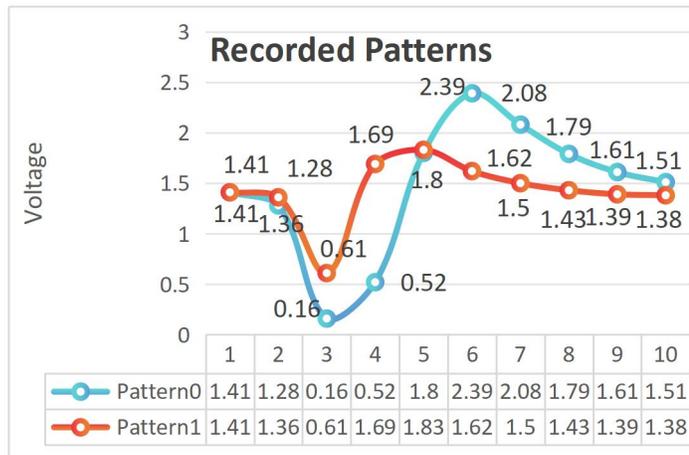

**Fig. 8.** Example recording of muscle patterns from Fig. 7.

These patterns are then searched through the live EMG signals of the user and every time one of the patterns is recognized by the DTW algorithm, the algorithm generates a "1" or a "0". Combining "1" s and "0"'s, with a small delay in between, users can generate multiple different motion patterns with just a single EMG.

*B. Accuracy and Statistics*

In the experiment below in Table 2, four different types of signals are tested in a confusion matrix. In four patterns, made out of the signals in Figure 7 above, the user can generate 4 combinations of "00", "01", "10" and "11". The confusion matrix calculates the accuracy of the algorithms for generating four patterns.

TABLE II
CONFUSION MATRIX OF INTENDED PATTERNS.

| | | Predicted | | | |
|---|---|---|---|---|---|
| | | 00 | 01 | 10 | 11 |
| Actual | 00 | 13 | | | |
| | 01 | | 11 | | |
| | 10 | | 1 | 9 | |
| | 11 | | | 3 | 13 |

After 46 successes and 4 failures, the algorithm provides a 92% accuracy, compared to other research on this subject, with 10 different patterns unlike 4, Castro et. al. [17] achieve 80% average accuracy. However, Castro's accuracy goes up to 97% as they reduce the patterns to 6. It is important to point out that Castro's calculations are done by 10 EMG sensors, unlike this study's single EMG.

Measured signals tend to provide delays between 414ms to 644ms between the signal applying and the motors start moving as shown in Table 3 below. This delay is due to the nature of Dynamic Time Warping making sure the single channel EMG signal is similar enough to the recorded signal pattern 1 and different enough then the pattern 0. In Table 3 below, the signal measurements are taken with a 23 ms intervals. The default state of the muscles provides a 1.85~ similarity signal to the searched pattern, when the similarity starts changing, it indicates the contraction is beginning. When the similarity value reaches below a certain number, in this case the value 1.12. The signal is determined to be similar enough to the searched pattern and motor starts to move the fingers with the intended outcome. In this case, the given delay is measured with the point where the signal is received, "14930ms" and, where the motor starts moving "15321ms", which gives a delay of "609 ms".

TABLE III
DELAY DUE TO THE NATURE OF DTW ALGORITHM

| Time (ms) | Signal Received | Action Taken | Time (ms) | Signal Received | Action Taken |
|---|---|---|---|---|---|
| 14700 | 1.84 | ☒ | 15022 | 1.19 | ☒ |
| 14723 | 1.89 | ☒ | 15045 | 1.19 | ☒ |
| 14746 | 1.89 | ☒ | 15068 | 1.19 | ☒ |
| 14769 | 1.89 | ☒ | 15091 | 1.19 | ☒ |
| 14792 | 1.89 | ☒ | 15114 | 1.19 | ☒ |
| 14815 | 1.89 | ☒ | 15137 | 1.22 | ☒ |
| 14838 | 1.89 | ☒ | 15160 | 1.22 | ☒ |
| 14861 | 1.89 | ☒ | 15183 | 1.22 | ☒ |
| 14884 | 1.89 | ☒ | 15206 | 1.22 | ☒ |
| 14907 | 1.89 | ☒ | 15229 | 1.22 | ☒ |
| 14930 | 1.19 | ☒ | 15252 | 1.22 | ☒ |
| 14953 | 1.19 | ☒ | 15275 | 1.22 | ☒ |
| 14976 | 1.19 | ☒ | 15298 | 1.22 | ☒ |
| 14999 | 1.19 | ☒ | 15321 | 1.12 | ☑ |

IV. DISSCUSSION

This paper establishes that a single EMG sensor with short impulse signals, along with Dynamic Time Warping (DTW) for pattern recognition, can be used to efficiently control a myoelectric prosthetic hand. This technique builds on top of the conventional multi-sensor methods by reducing the hardware specifications without sacrificing control precision.

The necessity of improving prosthetic technologies is highlighted by the crisis situations where limb injuries are common. The problems of existing multi-sensor myoelectric prosthetics (such as their expense, complexity, and lack of user satisfaction) are discussed in the introduction, which also highlights the need for a more straightforward and effective alternative.

This work is comparable to previous studies in the field that aim to enhance myoelectric prosthesis functionality and control. Similar to the work of Wattanasiri et al. [9] on multifunctional grip mechanisms and Atasoy et al. [8] on the development of a prosthetic hand with many degrees of freedom, the goal of this study is to improve control mechanisms to improve user experience. This study maintains the domain of EMG sensors, as opposed to Ahmed et al.'s [10] usage of pressure sensors, but it reduces hardware complexity and expense by simplifying the sensor arrangement to a single sensor.

This research leaves several unsolved questions. Although the study shows that a single EMG sensor can be used for control, it does not completely address how this strategy works in a variety of real-world settings, such as long-term use or varying muscle states. Furthermore, the influence of various forms of muscular tiredness on the precision of the control and the accuracy of the signal is not investigated. The comparison between this approach and more advanced multi-sensor systems in extremely dynamic and changing situations is another unanswered point in the study.

Subsequent investigations can build upon this work by verifying the robustness of the single-sensor technique in a wider range of demanding real-world scenarios. It is also vital to look into the system's long-term dependability and user adaptability. To improve the user experience, future development might concentrate on including more feedback systems, similar to haptic feedback with a piezo patch (a band that contracts with voltage) or a vibration motor (a motor that vibrates with voltage). A deeper understanding of how to best optimize prosthesis control may be gained from comparative research using alternative pattern recognition algorithms and sensor setups. Its influence and usefulness may also be increased by investigating how scalable this method is for various prosthetic devices and application kinds. In theory, the patterns that can be generated by simple binary pattern recognition can be limitless.

## V. Conclusion

Using Dynamic Time Warping (DTW) for pattern recognition, this study offers an improved method of controlling myoelectric prosthetic hands with a single EMG sensor and brief impulse signals. This approach represents a different perspective from conventional multi-sensor systems and shows that complex myoelectric control can be accomplished with less complicated gear. The findings show that the accuracy of the single-sensor, impulse-based method can effectively equal that of traditional multi-sensor systems. It may be possible to lower expenses, increase accessibility, and simplify the design of prosthetic devices through this simplification.

The study highlights the practicality of employing a low-tech sensor technique to enable natural and dependable control over prosthetic hands. This is particularly relevant in situations where advanced prostheses are desperately needed, including disaster recovery situations. This work creates new opportunities for the development of user-friendly prosthetic devices that can improve amputees' quality of life by concentrating on a less complicated and more affordable alternative.

Although the study yielded encouraging results, it also identified areas that require additional research, such as the system's long-term stability and its effectiveness in various real-world circumstances. In the end, this research adds to the enhancement of prosthetic technologies to offer more useful and accessible options for individuals requiring them.